\documentstyle[prl,aps,epsf,floats]{revtex}
\begin{document}
\draft
\twocolumn[\hsize\textwidth\columnwidth\hsize\csname @twocolumnfalse\endcsname
\title{Precursors of Antiferromagnetic and   
       Hubbard Bands  \\
       in the 2D Hubbard Model}
\author{Bumsoo Kyung}
\address{Max Planck Institute for Physics of Complex Systems, 
         Noethnitzer Str. 38, 01187 Dresden, Germany} 
\date{5 December 1997}
\maketitle
\begin{abstract}

  We formulate a theory to the 2D 
Hubbard model 
in a framework free of finite 
size effect and numerical analytical continuation, yet 
containing the essential features of the 2D Hubbard model, i.e., 
the correct atomic limit for large frequencies and 2D spin fluctuations.
As temperature is decreased for a 2D half-filled band,  
2D critical fluctuations give rise to 
a strong local maximum in $\mid Im \Sigma(\vec{k}_{F},\omega) \mid$
at $\omega=0$ leading to a split peak in the spectral function.
As $U$ is increased, four peaks associated with antiferromagnetic
and Hubbard bands begin to develop in small and intermediate
frequency regimes.  
\end{abstract}
\pacs{PACS numbers: 71.10.Fd, 71.27.+a}
\vskip2pc]
\narrowtext

   Recently the Hubbard model has received considerable attention,
since it is believed to contain the essential physics of 
strong electron correlations found in the high temperature 
superconductors\cite{Anderson:1987}.
Although it was solved exactly in one-dimension\cite{Lieb:1968},
the exact solution in higher dimensions is not known yet. 
For a half-filled 2D band, the $T=0$ ground state is believed to be 
antiferromagnetic insulating for all $U$, while at finite temperatures
strong spin fluctuations destroy long-range 
antiferromagnetic order 
due to the Mermin-Wagner theorem\cite{Mermin:1966}.

   In the absence of a small expansion parameter of the model in 
the physically relevant regime,
quantum Monte Carlo (QMC) simulations have played an important role 
in elucidating various dynamical properties at finite temperatures%
\cite{Dagotto:1994}.
In spite of its exact nature, QMC calculations suffer from two  
serious limitations, namely, small lattice size and numerical 
analytical continuation from imaginary frequencies to real ones.
In fact, 
depending on lattice size and on the method used to extract
$A(\vec{k},\omega)$ from  
$G(\vec{k},i\omega_{n})$ 
computed by QMC,
conflicting results have been reported 
for the single particle 
spectral function along the antiferromagnetic zone boundary 
for a half-filled 2D band with
$U=4t$ and $T=0.1-0.25t$%
\cite{White:1991,Creffield:1995}.
Although calculations based on the lowest order diagram (second order
in $U$) show the correct atomic limit for large $\omega$,
they fail to describe the correct 2D spin fluctuations.
This atomic limit is reflected as 
the correct asymptotic behavior
of the self-energy at large frequencies,
$U^{2}n/2(1-n/2)/i\omega_{n}$ (in the paramagnetic state),
which is necessary to produce the Hubbard bands at large enough $U$.
Recently
fluctuation exchange (FLEX) approximation 
was applied to the 2D Hubbard model  
by Bickers {\em et al.}\cite{Bickers:1989} and  
later extensively used by Bickers {\em et al.}\cite{Bickers:1991}
and by Dahm {\em et al.}\cite{Dahm:1995}.
Despite its successful application to the D-wave superconductivity 
and many other issues, it was pointed out 
by Vilk and Tremblay%
\cite{Vilk:1997}
that FLEX approximation does not show both the precursors of 
antiferromagnetic
and Hubbard bands
resulting from the correct 2D spin fluctuations and 
atomic limit for large $\omega$, respectively.
This is because FLEX approximation completely neglects the frequency dependent
vertex corrections for the self-energy within its self-consistent 
structure.
While the 2D spin fluctuations 
were quite successfully 
described by Vilk and Tremblay\cite{Vilk:1997}
by imposing the exact sumrules on the spin and charge susceptibilities,
their self-energy does not show the correct
asymptotic behavior at large frequencies.
Therefore,
it is highly desirable to formulate an approximation scheme for the 2D 
Hubbard model in a manner free of finite 
size effect and numerical analytical continuation, yet 
containing the essential features of the 2D Hubbard model, i.e.,
the correct atomic limit for large $\omega$ and 2D spin fluctuations.
In this Letter, we propose such a theory for the first time.

   We begin with reconsideration of paramagnon theory first proposed 
by Berk and Schrieffer\cite{Berk:1966}. 
This theory has 
some good features, namely, explicitly
enforcing the rotational invariance of the spin susceptibilities together
with including the particle-particle channel, and 
the correct prediction of the zero temperature phase 
transition for any $U$
in the half-filled 2D band.
This theory, however, always gives rise to 
a finite temperature antiferromagnetic instability
for large enough $U$ due to the insufficient treatment of strong   
2D spin fluctuations,
in conflict with the Mermin-Wagner theorem.
At frequencies larger than the bandwidth $W=8$, 
the self-energy obtained from this theory
is not guaranteed to give the correct  
asymptotic behavior.
Recently we found that 
within the same structure for the self-energy
and various susceptibilities as that in the paramagnon theory,
both the Mermin-Wagner 
theorem and correct atomic limit for large $\omega$
can be satisfied 
simultaneously 
by introducing  
renormalized interaction constants 
$U_{sp}$, $U_{ch}$, and $U_{pp}$  
in the spin, charge, and particle-particle channels, respectively.

   In our formulation the self-energy is expressed  
in terms of 
these renormalized interaction strengths to be determined later:
\begin{eqnarray}
 \Sigma(k) & = & \frac{U^{2}T}{N}\sum_{q} 
              \Bigg\{\bigg[ \chi^{0}_{ph}(q)
                 + \frac{3}{2}\chi^{0}_{ph}(q)
                   \Big( \frac{1}{1-U_{sp}\chi^{0}_{ph}(q)}-1
                   \Big)
                                                         \nonumber  \\
             & + & \frac{1}{2}\chi^{0}_{ph}(q)
                   \Big( \frac{1}{1+U_{ch}\chi^{0}_{ph}(q)}-1
                   \Big)
            \bigg]G^{0}(k-q)
                                                         \nonumber  \\
             & - & \bigg[ \chi^{0}_{pp}(q)
                   \Big( \frac{1}{1+U_{pp}\chi^{0}_{pp}(q)}-1
                   \Big)
            \bigg]G^{0}(q-k)\Bigg\}
                                               \; .
                                                           \label{eq1}
\end{eqnarray}
$T$, $N$ and $U$ are the absolute temperature, 
number of lattice sites and Coulomb repulsion energy.
$k$ is a compact notation for $(\vec{k},i\omega_{n})$ where
$i\omega_{n}$ are either Fermionic or 
Bosonic Matsubara frequencies. 
$G^{0}(k)$ is the noninteracting Green's function and 
$\chi^{0}_{ph}(q)$,
$\chi^{0}_{pp}(q)$ are 
irreducible particle-hole and particle-particle susceptibilities, 
respectively, which are computed from
\begin{eqnarray}
\chi^{0}_{ph}(q) & = & - \frac{T}{N}\sum_{k}G^{0}(k-q)G^{0}(k)
                                                         \nonumber  \\
\chi^{0}_{pp}(q) & = &  \frac{T}{N}\sum_{k}G^{0}(q-k)G^{0}(k)
                                               \; .
                                                           \label{eq2}
\end{eqnarray}
Dynamical spin, charge and particle-particle susceptibilities are 
calculated by 
\begin{eqnarray}
\chi_{sp}(q)&=&\frac{2\chi^{0}_{ph}(q)}{1-U_{sp}\chi^{0}_{ph}(q)}
                                                         \nonumber  \\
\chi_{ch}(q)&=&\frac{2\chi^{0}_{ph}(q)}{1+U_{ch}\chi^{0}_{ph}(q)}
                                                         \nonumber  \\
\chi_{pp}(q)&=&\frac{ \chi^{0}_{pp}(q)}{1+U_{pp}\chi^{0}_{pp}(q)}
                                               \; .
                                                           \label{eq3}
\end{eqnarray}
We determine
$U_{sp}$, $U_{ch}$, and $U_{pp}$  
by imposing the following three exact sumrules to Eq.~(\ref{eq3}):
\begin{eqnarray}
\frac{T}{N}\sum_{q}\chi_{sp}(q) & = & n-2\langle n_{\uparrow}n_{\downarrow}
                                         \rangle
                                                         \nonumber  \\
\frac{T}{N}\sum_{q}\chi_{ch}(q) & = & n+2\langle n_{\uparrow}n_{\downarrow}
                                         \rangle-n^{2}
                                                         \nonumber  \\
\frac{T}{N}\sum_{q}\chi_{pp}(q) & = & \langle n_{\uparrow}n_{\downarrow}
                                         \rangle
                                               \; ,
                                                           \label{eq5}
\end{eqnarray}
where the Pauli exclusion principle 
$\langle n^{2}_{\sigma} \rangle = \langle n_{\sigma} \rangle$  
is explicitly used in the spin and charge 
channels.
By finding  
$U_{sp}$, $U_{ch}$ and $U_{pp}$ 
through
Eq.~(\ref{eq5}), any possible magnetic instability  can happen only at 
zero temperature, because the right hand sides are always finite, 
in consistency with the Mermin-Wagner theorem.
The first two sumrules in Eq.~\ref{eq5} were previously used
by Vilk and Tremblay%
\cite{Vilk:1994}
to study collective spin and charge fluctuations in the 2D Hubbard model.
The correct asymptotic behavior of the self-energy
at large frequencies 
can be easily checked 
by substituting Eq.~(\ref{eq5}) and 
$\langle n_{\uparrow}n_{\downarrow} \rangle=n^{2}/4$ for 
noninteracting electrons into Eq.~(\ref{eq1}).
By making an ansatz
$U_{sp} \equiv U\langle n_{\uparrow}n_{\downarrow} \rangle/
(\langle n_{\uparrow} \rangle
\langle n_{\downarrow} \rangle)$\cite{Vilk:1994},
the double occupancy is computed self-consistently in the first 
sumrule in Eq.~(\ref{eq5})  
and $U_{ch}$, $U_{pp}$ can be obtained 
by the other two sumrules.
Throughout the calculations the unit of energy is $t$.
We used a $128 \times 128$ lattice in momentum space 
and performed the calculations by means of 
well-established fast Fourier transforms
(FFT).
It should be also noted that we used a real frequency formulation
in Eqs.~(\ref{eq1})-(\ref{eq5}) to avoid any possible uncertainties
associated with numerical analytical continuation.

\begin{figure}
 \vbox to 6.5cm {\vss\hbox to -5.0cm
 {\hss\
       {\includegraphics{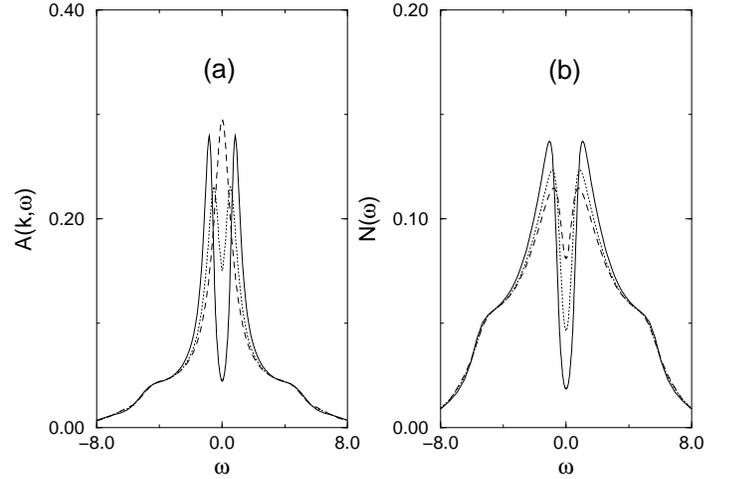}
       }
  \hss}
 }
\caption{(a) Spectral functions at $\vec{k}=(\pi/2,\pi/2)$ and 
         (b) density of states, for $U=4$ at
         $T=$0.10, 0.15 and 0.20 denoted as the solid, dotted and 
         dashed curves, respectively.}
\label{fig1}
\end{figure}

   As a first application of the present theory to the 2D Hubbard 
model, we study the half-filled 2D Hubbard model in the intermediate 
coupling regime ($U=4$) where 
QMC calculations have shown conflicting results 
for the single particle 
spectral function at the noninteracting Fermi surface%
\cite{White:1991,Creffield:1995}.
In this Letter we would like to give a definitive answer to this 
issue for the first time.
The spectral functions at $\vec{k}=(\pi/2,\pi/2)$ and 
the density of states
are presented
in Fig.~\ref{fig1}(a) and (b) for $U=4$ at
$T=0.1$, 0.15, and 0.2.  
At $T=0.2$ (dashed curve in Fig.~\ref{fig1}(a))
a single quasiparticle peak is found 
at the Fermi energy.
As the temperature is slightly decreased to $T=0.15$ (dotted curve), 
the single particle 
peak begins to split into two, 
leading to a pseudogap%
\cite{Kampf:1990}  
at the Fermi energy.
This is a 
precursor of 
antiferromagnetic bands resulting from strong 2D spin     
fluctuations at low temperatures.    
At $T=0.1$ (solid curve), the two peak structure with a pseudogap inside
becomes much more pronounced. 
The present results support the existence of a pseudogap 
in the spectral function at $T < 0.2$
for $U=4$.
For all three temperatures, however, the density of 
states (Fig.~\ref{fig1}(b)) shows a pseudogap at the Fermi energy.
In spite of a single peak structure in the spectral function
for $T=0.2$ (dashed curve),
the density of states is significantly suppressed 
at the Fermi energy compared with that for non-interacting electrons
which shows a logarithmic divergence.
Because of its accumulative nature, the density of states appears more 
sensitive to the change of a quasiparticle state at the Fermi energy
than the spectral function itself does, at least, for a half-filled 2D band.
This situation for $T=0.2$ is the best result
found by Dahm {\em et al.}\cite{Dahm:1995}
within FLEX approximation in an effort to see the precursor
effect of the 
antiferromagnetic bands.    
The small bumps at $\pm (4-5)t$ both in the spectral function and 
density of states are precursors of the Hubbard bands, which will 
be discussed later.

\begin{figure}
 \vbox to 6.5cm {\vss\hbox to -5.0cm
 {\hss\
       {\includegraphics{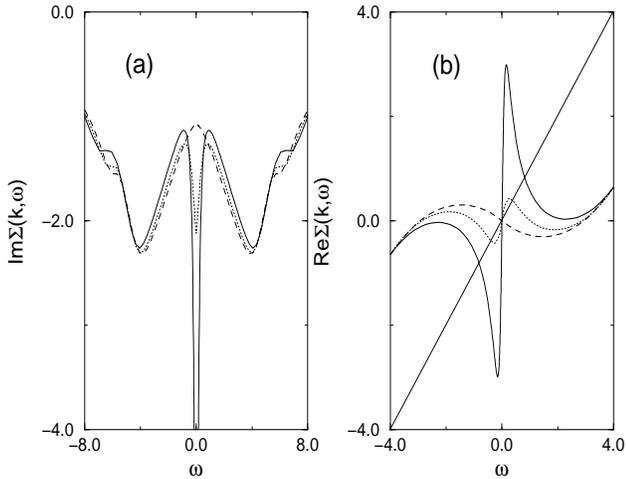}
       }
  \hss}
 }
\caption{(a) Imaginary and (b) real parts of the self-energy  
         at $\vec{k}=(\pi/2,\pi/2)$ for $U=4$ at 
         $T=$0.1, 0.15 and 0.2 denoted as the solid, dotted and 
         dashed curves, respectively.
         The solid straight line in (b) is $\omega-\varepsilon(\vec{k})$.}
\label{fig2}
\end{figure}

   This anomalous behavior is understood more clearly by examining 
the imaginary and real parts of the self-energy shown
in Fig.~\ref{fig2}(a) and Fig.~\ref{fig2}(b), respectively 
for the same parameters as 
in Fig.~\ref{fig1}.  
At $T=0.2$ (dashed curve),
the imaginary part of the self-energy becomes smaller in magnitude
as the Fermi energy is approached, indicating a Fermi liquid-like
behavior. 
As the temperature is further decreased to $T=0.15$ (dotted curve)
to $T=0.1$ (solid curve), however,
a drastic change happens near the Fermi energy.
Due to 2D critical fluctuations, the scattering rates
at the Fermi energy grow exponentially as 
$\sim \xi \sim exp(constant/T)$. 
This singular scattering by exchange of strong 2D
spin fluctuations is 
responsible for the strong suppression of the spectral weight at the 
Fermi energy, leading to the formation of a pseudogap in the spectral
function.

  The corresponding real part of the self-energy 
obtained from the imaginary part by means of 
the Kramers-Kronig relations
is shown in Fig.~\ref{fig2}(b).
The peak condition in the spectral function
is determined by the intersection of the real part of 
the self-energy and 
$\omega=\varepsilon(\vec{k})$ denoted as the solid  
straight line in Fig.~\ref{fig2}(b). 
$\varepsilon(\vec{k})=-2t(\cos k_{x} + \cos k_{y})$ for nearest   
neighbor hopping.
At $T=0.2$ (dashed curve), the peak condition is satisfied 
only at the Fermi energy with its slope 
being negative, a characteristic feature of the 
Fermi liquid.
As the temperature is decreased to $T=0.15$ (dotted curve)
to $T=0.1$ (solid curve),  
its slope  
becomes positive and larger than unity at the Fermi energy, 
and thus the peak condition is satisfied at 
three different locations.
Because of large scattering rates at the Fermi energy, however, only two peaks
appear in the spectral function,
consistent with the results
in Fig.~\ref{fig1}.
\begin{figure}
 \vbox to 6.5cm {\vss\hbox to -5.0cm
 {\hss\
       {\includegraphics{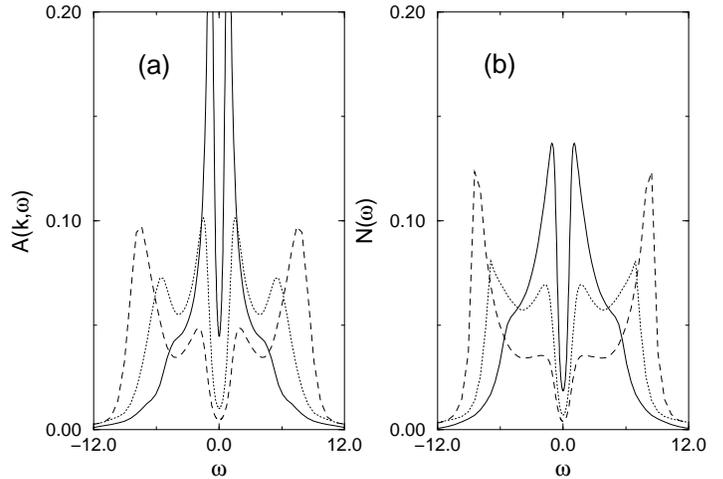}
       }
  \hss}
 }
\caption{(a) Spectral functions at $\vec{k}=(\pi/2,\pi/2)$ and 
         (b) density of states, for
         $U=$4, 6 and 8 at $T=0.1$ denoted as the solid, dotted and 
         dashed curves, respectively.}
\label{fig3}
\end{figure}

   Since the present formulation for the 2D Hubbard model can describe
quite reasonably 
both 2D critical spin fluctuations 
and the correct atomic limit for large $\omega$,
it is of 
great interest to examine
how the precursors of  
antiferromagnetic and Hubbard bands    
evolve as the interaction strength is increased. 
For $U=4$ (solid curve in Fig.~\ref{fig3}(a)),  
a split peak 
with developing small bumps 
at $\pm (4-5)t$ is found.
As $U$ is increased to 6 (dashed curve), the  
antiferromagnetic bands become  
significantly suppressed 
and the Hubbard bands grow substantially near $\pm 6$.
For $U=8$ (solid curve),
the spectral weight inside the Hubbard bands becomes
further suppressed due to the large Coulomb repulsion, and 
as a result 
the spectral weight associated with the Hubbard bands
becomes dominating over that with the antiferromagnetic bands.  
The Hubbard bands for $U=8$ occur at larger frequencies than $\pm U/2$, since 
the asymptotic behavior
of the self-energy,
$U^{2}n/2(1-n/2)/i\omega_{n}$, 
sets in at much higher frequencies than the bandwidth. 
This is 
because the strong maximum of the 
scattering rates at the Fermi energy 
significantly enhances the real part up to high frequencies.
The appearance of four peaks in the spectral function for $U=8$ is 
consistent with recent
QMC calculations%
\cite{Moreo:1995,Preuss:1995} except a difference in the relative 
strength of the two different bands.
This feature is also similar to the one from the 1/d effect in
the infinite dimensional Falicov-Kimball model\cite{Laad:1997}.
The density of states obtained from the 
average of the spectral function in the whole Brillouin zone
shows more suppressed antiferromagnetic bands
and thus more pronounced Hubbard bands 
in Fig.~\ref{fig3}(b)  
than the spectral  
function does at $(\pi/2,\pi/2)$.

\begin{figure}
 \vbox to 6.5cm {\vss\hbox to -5.0cm
 {\hss\
       {\includegraphics{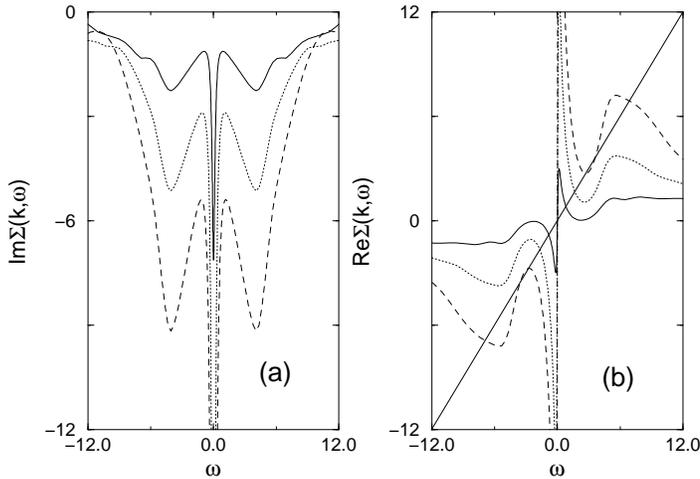}
       }
  \hss}
 }
\caption{(a) Imaginary and (b) real parts of the self-energy  
         at $\vec{k}=(\pi/2,\pi/2)$ for 
         $U=$4, 6 and 8 at $T=0.1$ denoted as the solid, dotted and 
         dashed curves, respectively.
         The solid straight line in (b) is $\omega-\varepsilon(\vec{k})$.}
\label{fig4}
\end{figure}

   As the interaction strength $U$ is increased,
the imaginary part of the self-energy grows rapidly 
both at the Fermi energy and in the intermediate frequency
region ($\sim \pm 4$) shown 
in Fig.~\ref{fig4}(a).
The former comes from 
strong 2D critical fluctuations near half-filling and the latter
mainly from the lowest order diagram (second order in $U$)
which is responsible for the Hubbard bands for large enough $U$.
As $U$ is increased to $8$ (dashed curve in Fig.~\ref{fig4}(b)),
seven solutions are found in the peak condition.
The most outer two solutions lead to the Hubbard bands and 
the most inner two near the Fermi energy to
the antiferromagnetic bands, and the other three accompanied by 
large scattering rates yield the incoherent 
background.
This result is similar to the earlier report by 
Kampf and Schrieffer\cite{Kampf:1990} of the developing multiple 
solutions in the peak condition 
by increasing the spin-spin correlation length.
Besides the total number of solutions, however,
there are some qualitative differences between these two results. 
Because of the insufficient treatment of the 2D critical fluctuations
in their phenomenological 
antiferromagnetic spin fluctuation spectrum,
their solution at the Fermi energy shows a single quasiparticle peak,
while in our calculations a quasiparticle state is destroyed for  
a half-filled band.
Due to the same reason,
the pseudogap found in their paper does not come from a
suppressed spectral weight inside the {\it antiferromagnetic} bands  
but instead inside 
the {\it Hubbard} bands.

  In summary, we formulated a theory to the 2D 
Hubbard model in a manner free of finite 
size effect and numerical analytical continuation, yet 
containing the essential features of the 2D Hubbard model, i.e.,
the correct atomic limit for large $\omega$ and 2D spin fluctuations.
As the temperature is decreased for a 2D half-filled band,  
anomalous behaviors 
are found near the Fermi energy
such as a split peak in the spectral function,
a large positive slope greater than unity
in the real part of the self-energy and 
a strong local maximum in the scattering rates. 
As the interaction strength $U$ is increased, four peaks associated with
the antiferromagnetic and Hubbard bands begin to develop 
in the small and intermediate
frequency regimes.  

    The author would like to thank Prof. P. Fulde for critical reading 
of the manuscript as well as 
for his valuable comments.
Useful discussions with Drs. S. Blawid, R. Bulla, T. Dahm,  
P. Kornilovitch, M. Laad, W. Stephan, and numerous other colleagues 
in the Max Planck Institute
for Physics of Complex Systems would like to be 
acknowledged.

\begin{references}
%
%
\bibitem{Anderson:1987} P. W. Anderson, Science {\bf 235},
                       1196 (1987).
\bibitem{Lieb:1968} E. H. Lieb and F. Y. Wu, Phys. Rev. Lett. {\bf 20},
                       1445 (1968).
\bibitem{Mermin:1966} N. D. Mermin and H. Wagner, Phys. Rev. Lett. {\bf 17},
                       1133 (1966).
\bibitem{Dagotto:1994} E. Dagotto, Rev. Mod. Phys. {\bf 66},
                       763 (1994).
\bibitem{White:1991} S. R. White, Phys. Rev. B {\bf 44},
                       4670 (1991); M. Veki\'{c} and S. R. White,
                       {\it ibid.} {\bf 47} 1160 (1993).
\bibitem{Creffield:1995} C. E. Creffield {\it et al.},
                       Phys. Rev. Lett. {\bf 75},
                       517 (1995).
\bibitem{Bickers:1989} N. E. Bickers, D. J. Scalapino, and S. R. White,
                       Phys. Rev. Lett. {\bf 62},
                       961 (1989).
\bibitem{Bickers:1991} N. E. Bickers and S. R. White,
                       Phys. Rev. B {\bf 43},
                       8044 (1991).
\bibitem{Dahm:1995} T. Dahm and L. Tewordt,
                       Phys. Rev. B {\bf 52},
                       1297 (1995);
                       J. J. Deisz, D. W. Hess, and J. W. Serene,
                       Phys. Rev. Lett. {\bf 76},
                       1312 (1996).
\bibitem{Vilk:1997} Y. Vilk and A. M. Tremblay, to appear in J. Physics
                       (Paris)(Nov.1997); cond-mat/9702188.
\bibitem{Berk:1966} N. F. Berk and J. R. Schrieffer,
                       Phys. Rev. Lett. {\bf 17},
                       433 (1966);
                    P. C. E. Stamp, J. Phys. F: Met. Phys.,
                       {\bf 15}, 1829 (1985).
\bibitem{Vilk:1994} Y. M. Vilk, Liang Chen, and A. M. Tremblay,
                       Phys. Rev. B {\bf 49},
                       13 267 (1994). 
\bibitem{Kampf:1990} A. Kampf and J. R. Schrieffer, Phys. Rev. B {\bf 41},
                       6399 (1990); {\it ibid.} {\bf 42} 7967 (1990).
\bibitem{Moreo:1995} A. Moreo {\it et al.},
                       Phys. Rev. B {\bf 51},
                       12 045 (1995).
\bibitem{Preuss:1995} R. Preuss, W. Hanke, and W. von der Linden,
                       Phys. Rev. Lett. {\bf 75},
                       1344 (1995).
\bibitem{Laad:1997} M. Laad and M. Van den Bossche (unpublished).
%
%
\end{references}
\end{document}